\documentclass{article}
\usepackage{amsmath, amssymb, amsfonts}
\title{Semiclassical corrections to a regularized Schwarzschild metric \footnote{Presented at The 3rd Conference of the Polish Society on Relativity,  25-29 September 2016, Kraków, Poland}}
\author{Hristu Culetu \footnote{e-mail : hculetu@yahoo.com} \\Ovidius University, Deparment of Physics \\ Bld. Mamaia 124, 900527  Constanta, Romania}

\begin{document}
\numberwithin{equation}{section}
\pagenumbering{arabic}
\maketitle
\newcommand{\fv}{\boldsymbol{f}}
\newcommand{\tv}{\boldsymbol{t}}
\newcommand{\gv}{\boldsymbol{g}}
\newcommand{\OV}{\boldsymbol{O}}
\newcommand{\wv}{\boldsymbol{w}}
\newcommand{\WV}{\boldsymbol{W}}
\newcommand{\NV}{\boldsymbol{N}}
\newcommand{\hv}{\boldsymbol{h}}
\newcommand{\yv}{\boldsymbol{y}}
\newcommand{\RE}{\textrm{Re}}
\newcommand{\IM}{\textrm{Im}}
\newcommand{\rot}{\textrm{rot}}
\newcommand{\dv}{\boldsymbol{d}}
\newcommand{\grad}{\textrm{grad}}
\newcommand{\Tr}{\textrm{Tr}}
\newcommand{\ua}{\uparrow}
\newcommand{\da}{\downarrow}
\newcommand{\ct}{\textrm{const}}
\newcommand{\xv}{\boldsymbol{x}}
\newcommand{\mv}{\boldsymbol{m}}
\newcommand{\rv}{\boldsymbol{r}}
\newcommand{\kv}{\boldsymbol{k}}
\newcommand{\VE}{\boldsymbol{V}}
\newcommand{\sv}{\boldsymbol{s}}
\newcommand{\RV}{\boldsymbol{R}}
\newcommand{\pv}{\boldsymbol{p}}
\newcommand{\PV}{\boldsymbol{P}}
\newcommand{\EV}{\boldsymbol{E}}
\newcommand{\DV}{\boldsymbol{D}}
\newcommand{\BV}{\boldsymbol{B}}
\newcommand{\HV}{\boldsymbol{H}}
\newcommand{\MV}{\boldsymbol{M}}
\newcommand{\be}{\begin{equation}}
\newcommand{\ee}{\end{equation}}
\newcommand{\ba}{\begin{eqnarray}}
\newcommand{\ea}{\end{eqnarray}}
\newcommand{\bq}{\begin{eqnarray*}}
\newcommand{\eq}{\end{eqnarray*}}
\newcommand{\pa}{\partial}
\newcommand{\f}{\frac}
\newcommand{\FV}{\boldsymbol{F}}
\newcommand{\ve}{\boldsymbol{v}}
\newcommand{\AV}{\boldsymbol{A}}
\newcommand{\jv}{\boldsymbol{j}}
\newcommand{\LV}{\boldsymbol{L}}
\newcommand{\SV}{\boldsymbol{S}}
\newcommand{\av}{\boldsymbol{a}}
\newcommand{\qv}{\boldsymbol{q}}
\newcommand{\QV}{\boldsymbol{Q}}
\newcommand{\ev}{\boldsymbol{e}}
\newcommand{\uv}{\boldsymbol{u}}
\newcommand{\KV}{\boldsymbol{K}}
\newcommand{\ro}{\boldsymbol{\rho}}
\newcommand{\si}{\boldsymbol{\sigma}}
\newcommand{\thv}{\boldsymbol{\theta}}
\newcommand{\bv}{\boldsymbol{b}}
\newcommand{\JV}{\boldsymbol{J}}
\newcommand{\nv}{\boldsymbol{n}}
\newcommand{\lv}{\boldsymbol{l}}
\newcommand{\om}{\boldsymbol{\omega}}
\newcommand{\Om}{\boldsymbol{\Omega}}
\newcommand{\Piv}{\boldsymbol{\Pi}}
\newcommand{\UV}{\boldsymbol{U}}
\newcommand{\iv}{\boldsymbol{i}}
\newcommand{\nuv}{\boldsymbol{\nu}}
\newcommand{\muv}{\boldsymbol{\mu}}
\newcommand{\lm}{\boldsymbol{\lambda}}
\newcommand{\Lm}{\boldsymbol{\Lambda}}
\newcommand{\opsi}{\overline{\psi}}
\renewcommand{\tan}{\textrm{tg}}
\renewcommand{\cot}{\textrm{ctg}}
\renewcommand{\sinh}{\textrm{sh}}
\renewcommand{\cosh}{\textrm{ch}}
\renewcommand{\tanh}{\textrm{th}}
\renewcommand{\coth}{\textrm{cth}}

\begin{abstract}
A regular form of the Schwarzschild geometry is proposed. It is more suitable for application in microphysics because the source mass comes out both as a Schwarzschild radius and the Compton wavelength of the mass $m$. The Komar energy equals $mc^{2}$ in theclassical situation ($\hbar = 0$).
\end{abstract}

We propose a regular version of the Schwarzschild metric, to be valid in microphysics. The time-time metric coefficient is modified as \cite{HC} (see also \cite{LX})
     \begin{equation}
   -g_{tt} = 1/g_{rr} \equiv f(r) = 1 - \frac{2m}{r} e^{-\frac{k}{mr}},   
 \label{1}
 \end{equation}  
where $m$ is the object mass, $k$ is a positive dimensionless constant and has units of length in front of the exponential and $1/length$ at the exponent. We select $k = 2/e$, so that $f(r)$ becomes minimal at $r = k/m = 2/me$. For a horizon to exist, we found that the condition $m \geq m_{P}$ should be obeyed \cite{BR}. 

An expansion of $f(r)$ for $r >> r_{0} = 2/em$ gives us
 \begin{equation}
   f(r) \approx 1 - \frac{2m}{r} + \frac{4l_{P}^{2}}{er^{2}},
 \label{2}
 \end{equation}  
where $l_{P}$ is the Planck length. From (0.2) one obtains that $f(r)$ acquires its Schwarzschild value when $\hbar = 0$. The solution (0.1) is not a vacuum solution of Einstein's equations. The source stress tensor has $p_{r} = -\rho$ and fluctuating transversal pressures, where $\rho$ is the energy density and $p_{r}$ is the radial pressure. The Komar energy associated to the geometry (0.1), with $k = 2/em$, appears as
 \begin{equation}
 W = \left(mc^{2} - \frac{2\hbar c}{er}\right) e^{-\frac{2\hbar}{emcr}}.
\label{3}
\end{equation}
which tends to zero when $r \rightarrow 0$ and $W \rightarrow mc^{2}$ at infinity. The classical situation ($\hbar = 0$) leads to the standard result $ W = mc^{2}$.

\end{document}